\documentclass[useAMS,usenatbib,usegraphicx]{mn2e}
\usepackage{journals}   % only needed locally to make .bbl, remove before submission

\begin{document}

\def\fracd#1#2{{\displaystyle\frac{#1}{#2}}}

\title[Implications of the Galactic aberration for CRF]
{On the implications of the Galactic aberration in proper motions for celestial reference frame}

\author[Z. Malkin]{Zinovy Malkin$^{1,2}$ \\ $^1$Pulkovo Observatory, St.~Petersburg 196140, Russia\\
  $^2$St. Petersburg State University, St.~Petersburg 198504, Russia}
\maketitle

\begin{abstract}
During the last years, much attention has been paid to the astrometric implications of the galactic aberration in proper motions (GA).
This effect causes systematic errors in astrometric measurements at a $\mu$as level.
Some authors consider it so serious that it requires redefinition of the celestial reference system (CRF).
We argue that such attention to the GA is too much exaggerated.
It is just a small astrometric correction that must be taken into account during highly accurate astrometric and geodetic data processing.
The accuracy of this correction depends on accuracy of the Galactic rotation parameters and, for most application, on the accuracy of the
rotation matrix between Galactic and equatorial systems.
Our analysis has shown that our today knowledge of these two factors is sufficient to compute the GA correction with accuracy of better than 10\%.
The remaining effect at level a few tenths $\mu$as/yr is negligible nowadays.
Another consequence of introducing the GA correction is necessity to return to classical astrometric modeling of the VLBI-derived extragalactic
radio source position by the linear trend model.
Changing the current paradigm of VLBI-derived CRF based on assumption of zero motion of radio sources to classical one leads to bias
in the radio source positions up to several tens $\mu$as for catalog at epoch J2000.0.
\end{abstract}


%%%%%%%%%%%%%%%%%%%%%%%%%%%%%%%%%%%%%%%%%%%%%%%%%%%%%%%%%%%%%%%%%%%%%%%%%%%%%%%%%

\section{Introduction}

Astronomical observations conducted from the Earth's surface or from another place in the Solar system are affected by systematic error in proper motions
caused by rotation of the Galaxy, more specifically by the Galactocentric acceleration of the Sun.
The terminology related to this effect is not settled yet, see discussion in \cite{Liu2012}.
We use hereafter the term Galactic aberration in proper motion (shortly GA) for consistency with our previous publications.
The GA is well studied theoretically by \citet{Kovalevsky2003,Kopeikin2006,Liu2012,Liu2013}.
The most evident effect coming from the GA is systematics in the apparent motions of celestial objects at a level of several $\mu$as.
Through this effect, the GA impacts most of key astrometric and geodetic products such as celestial reference frame (CRF) and Earth orientation
parameters (EOP) \citep{Titov2010,Malkin2011fe,Liu2012,Xu2013}.

In this paper, we focus on the implications of the GA for CRF, more specifically for the International Celestial Reference Frame (ICRF).
The first estimation of the ICRF rotation due to the GA was made by \citet{Liu2012}.
It has been shown that not taking the GA into account leads to slow rotation of the VLBI-derived CRF at the level of about 1~$\mu$as/yr.
The exact rotational effect cannot be predicted theoretically as it strongly depends on the CRF objects distribution over the sky.
In fact, it is different for each CRF catalog.
The more homogeneous is the source distribution over the sky, the less rotation effect is expected.
However the sources distribution in ICRF2 is far from uniform \citep{Ma2009p}.

Based at these and their own results, some authors propose redefinition of the ICRS and ICRF \citep{Titov2010,Xu2013}.
However, we believe they dramatize the implications of the GA for astrometric results.
We argue that the GA is just an astrometric reduction like annual aberration or precession, which can and should be including in routine procedures
of highly-accurate data processing, which would allow us to eliminate the consequences of this effect in the most natural way.

The main question is whether we can model the GA contribution with sufficient accuracy.
In this paper we analyze this problem and show that it quite possible.
We consider here only the observations of extragalactic celestial objects because the ICRF is based on these measurements.
The GA effect on the observations of the Galactic objects is more complicated and considered in detail by \citet{Kovalevsky2003,Liu2013}.

%%%%%%%%%%%%%%%%%%%%%%%%%%%%%%%%%%%%%%%%%%%%%%%%%%%%%%%%%%%%%%%%%%%%%%%%%%%%%%%%%%

\section{GA modeling}

The GA-induced proper motion is given by \citep{Kovalevsky2003,Malkin2011fe}:
\begin{equation}
\begin{array}{rcl}
\mu_l \cos b &=& -A \sin l \,, \\
\mu_b &=& -A \cos l \sin b \,, \\
\end{array}
\label{eq:galactic}
\end{equation}
where $l$ and $b$ are Galactic longitude and latitude of the celestial body, respectively, and $A$ is the GA constant, which depends on the Galactic
rotation parameters:
\begin{equation}
A = \frac{V_0 \Omega_0}{c} = \frac{R_0 \Omega_0^2}{c} = \frac{V_0^2}{R_0 \, c} \,,
\label{eq:Aconst}
\end{equation}
where $V_0$ and $\Omega_0$ are the linear and angular speed, respectively, of the circular motion of the local standard of rest (LSR) due to
the rotation of the Galaxy, $R_0$ is the Galactocentric distance of the Sun, and $c$ is the speed of light.

\begin{figure}
\centering
\includegraphics[width=0.48\textwidth,clip]{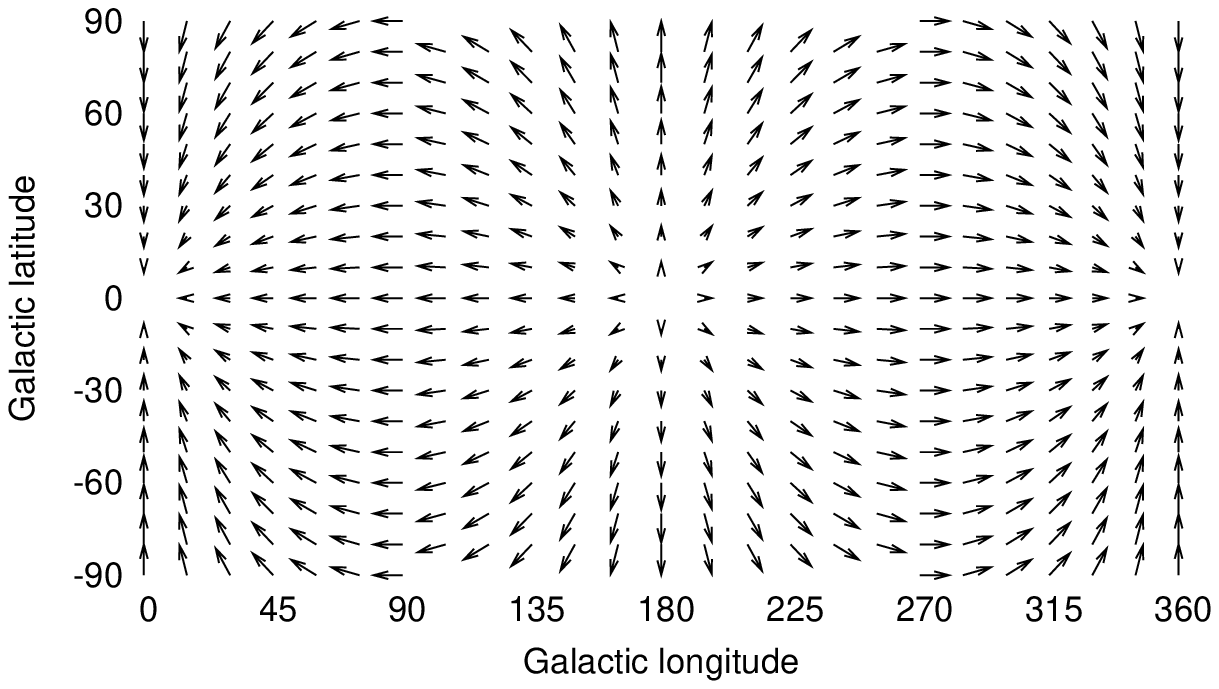}
\includegraphics[width=0.48\textwidth,clip]{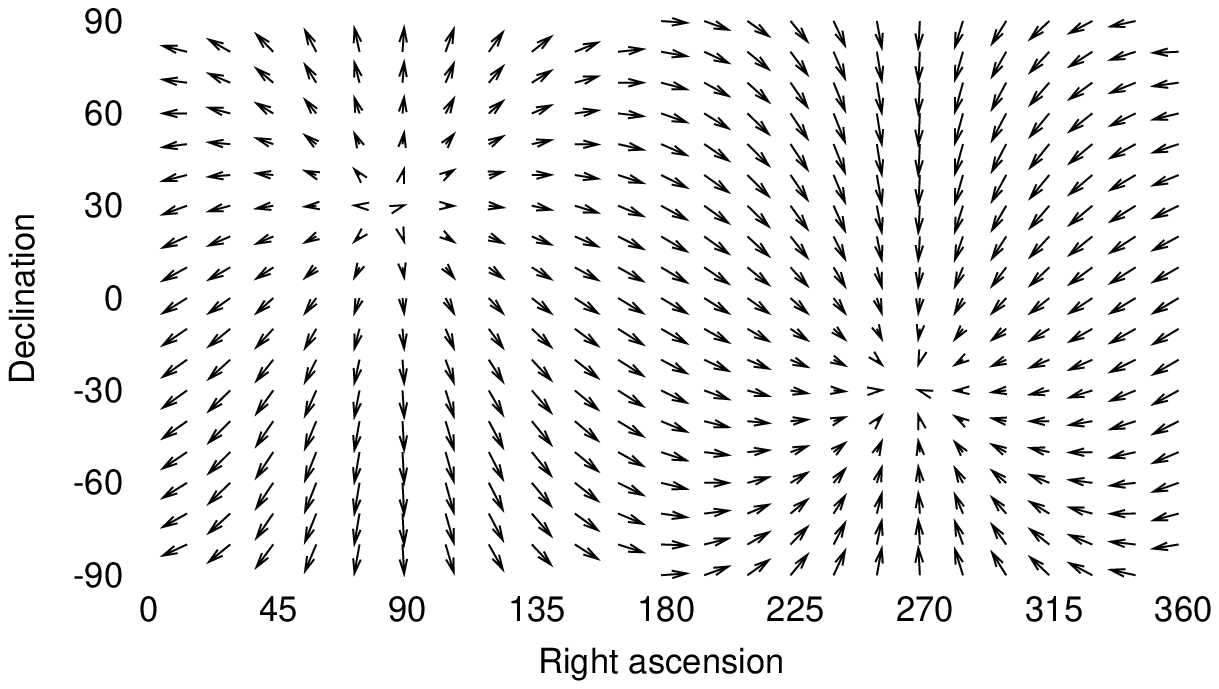}
\caption{Apparent motions caused by galactic aberration in galactic and equatorial coordinate systems.
The longest arrows correspond to motion of 5~$\mu$as/yr.
Note that $\mu_l$ are multiplied by $\cos b$ and $\mu_\alpha$ are multiplied by $\cos \delta$.}
\label{fig:pm_aberr}
\end{figure}

The distribution of the apparent proper motions caused by the GA over the celestial sphere is shown in Fig.~\ref{fig:pm_aberr}.
The amplitude of this motion in Galactic coordinate system is equal to $A \sqrt{\sin^2 l + \cos^2 l \sin^2 b}~\mu$as/yr,
which is minimum (zero) at the Galactic center and anticenter ($l=0,180^\circ, b=0$), and maximum ($A$) at $l=90,270^\circ$ independently of $b$.
Correspondingly, in equatorial coordinate system the GA effect vanishes at the points $(\alpha=86.4^\circ, \delta=+28.4^\circ)$, 
$(\alpha=266.4^\circ, \delta=-28.4^\circ)$ and has maximum magnitude at $\alpha=176.4, 356.4^\circ$.

An attempt was made in this study to estimate the actual rotation of VLBI-based CRF.
For this purpose we investigated the orientation between ICRF2 \citep{Ma2009p} and four catalogs of the radio source position computed at
the NASA Goddard Space Flight Center (GSFC) in 2010, 2011, 2012, and 2014 using a strategy similar to that used for computation
of ICRF2\footnote{http://gemini.gsfc.nasa.gov/solutions/astro/}.
The orientation between two CRF realizations (catalogs) is described by three angles $A_1$, $A_2$, and $A_3$ of rotation around three coordinate axes:
\begin{equation}
\begin{array}{rcl}
\Delta\alpha & = & \phantom{-}A_1\cos\alpha\tan\delta + A_2\sin\alpha\tan\delta - A_3 \,, \\
\Delta\delta & = & -A_1\sin\alpha + A_2\cos\alpha \,,
\end{array}
\label{eq:rotation_tanDE}
\end{equation}
where $\Delta\alpha, \Delta\delta$ are coordinates differences between two catalogs.

\begin{figure}
\centering
\includegraphics[clip,width=0.48\textwidth]{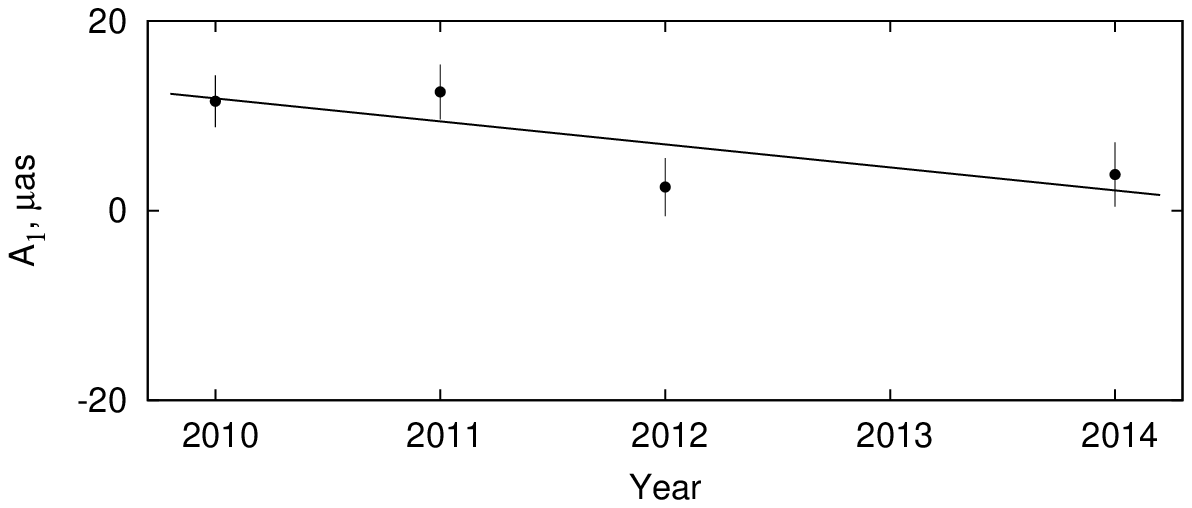}
\includegraphics[clip,width=0.48\textwidth]{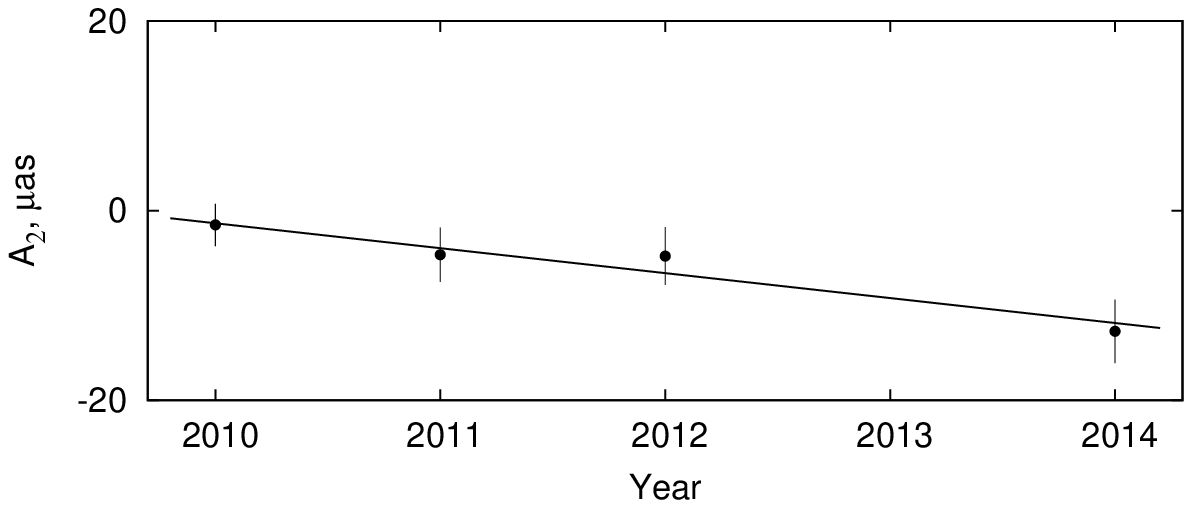}
\includegraphics[clip,width=0.48\textwidth]{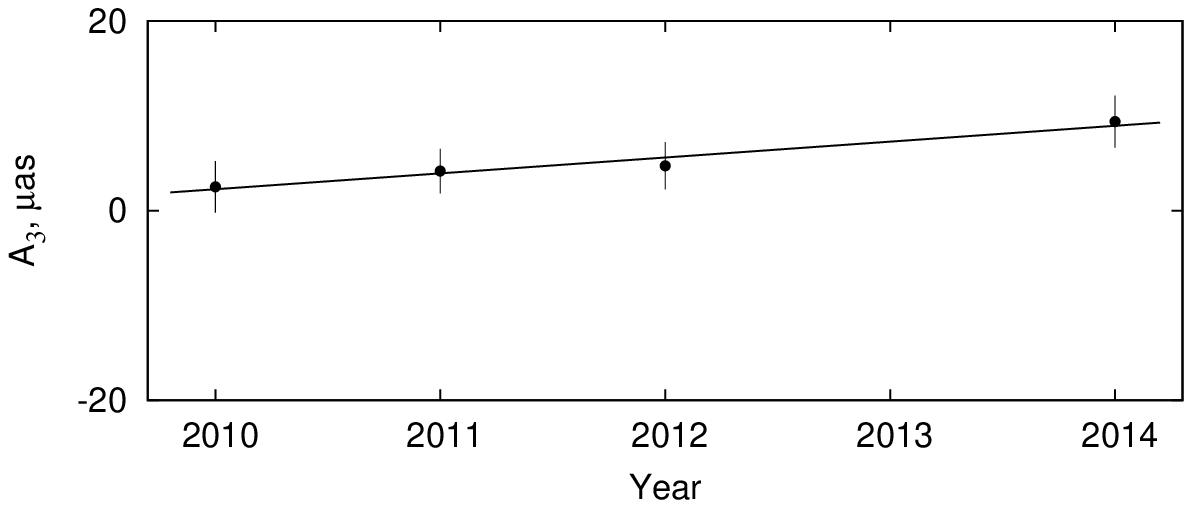}
\caption{Evolution with time of the orientation angles between ICRF2 and GSFC catalogs.}
\label{fig:orientation}
\end{figure}

Figure~\ref{fig:orientation} shows the evolution of these angles with time.
One can see a substantial graduate rotation of the GSFC catalogs.
The final interpretation of this rotation requires separate detailed discussion, but the GA effect is definitely one of the most probable reason.

Now we consider the accuracy of computation of the GA contribution to the astrometric reduction of the observations.

\subsection{GA constant}

Imperfect knowledge of the GA constant $A$ is the first source of the uncertainty in modeling the GA effect.
The $A$ value can be derived in two ways.

First, it can be computed directly by Eq.~\ref{eq:Aconst} using the values of the Galactic rotation parameters as measured by the methods of the
Galactic astronomy.
Tens of such determinations were made during last decades using different approaches.
Most promising results are obtained by means of the direct VLBI measurements of distances in the Galaxy.

Our first estimate of the GA constant made in \citet{Malkin2011fe} was based on the several last estimates of $R_0$ and $\Omega_0$, which gave
$A = 5.02~\mu$as/yr (uncertainty was not estimated).
Computation recently made in \citet{Malkin2014a} using all available measurements of the Galactic rotation parameters made during last 5 years
yields the result of $A = 5.0 \pm 0.3~\mu$as/yr.
This result practically does not depend on whether only direct or all available measurements of the Galactic rotation parameters are used.
Thus, it can be suggested as the current best estimate based on the stellar astronomy results.
Our result can be compared with the value $A = 4.9 \pm 0.4 \mu$as/yr derived from the latest estimates of the Galactic rotation constants
$R_0 = 8.34 \pm 0.16$ kpc and the circular rotation speed $V_0 = 240 \pm 8$ \citep{Reid2014}.
All these results are in very good agreement.

On the other hand, the duration and accuracy of astrometric VLBI time series is now sufficient to try to estimate $A$ directly
from systematic changes in the radio source positions.
Several analysis strategies can be used to estimate the Galactocentric LSR acceleration \citep{Malkin2009c}:
\begin{itemize}
\item[a)] estimate source position and velocities from global solution, then fit spherical harmonics to the velocities;
\item[b)] compute the coefficients of spherical harmonics as global parameters;
\item[c)] compute velocities from position time series, then fit spherical harmonics to the velocities.
\end{itemize}
In the fourth method proposed by \citet{Xu2012b}, the Galactocentric Sun acceleration is obtained by fitting the solar velocity variation time series.
All these methods were used in practice.
Consider the latest results obtained by different groups using different software and analysis strategy.

\citet{Kurdubov2011} estimated the Galactocentric acceleration of the Sun as global parameter and obtained the value
$a = (4.7 \pm 0.5) \cdot 10^{-10}~\mathrm{m/s^2}$, which is equivalent to $A = 10.2 \pm 1.1~\mu$as/yr.

\citet{Xu2012b} obtained two estimates of $a$.
First one was computed as a global parameter, which yields $a = 7.47 \pm 0.46$ mm/s/yr ($A = 5.1 \pm 0.3~\mu$as/yr).
Second estimate was obtained from the solar velocity time series solution, which give $a = 9.10 \pm 1.74$~mm/s/yr ($A = 6.3 \pm 1.2~\mu$as/yr).

\citet{Titov2013} obtained two estimates of $a$ from analysis of the source position time series:
$a = 9.3 \pm 1.1$ mm/s/yr ($A = 6.4~\pm 0.8~\mu$as/yr for DR solution (estimated dipole and rotation parameters) and
$a = 10.1 \pm 1.2$ mm/s/yr ($A = 7.8~\pm 0.8 \mu$as/yr for DRQ solution (estimated dipole, rotation, and the quadrupole parameters).
The authors also revealed strong dependence of the results on data editing.

Considering the results obtained from VLBI observations we can see that they are yet poorly consistent.
This problem was also discussed by \citet{Malkin2009c,Kurdubov2011,Titov2011}.
The results of measurements of the Galactic rotation parameters by the methods of stellar astronomy are much more robust.
Thus we can suggest the current best estimate of the GA constant $A = 5 \pm 0.3~\mu$as/yr for modeling of the GA effect on the source positions.
Using this value of the GA constant allows one to eliminate more than 90\% of the GA effect.
Remaining uncertainty in proper motion of less than 0.5~$\mu$as/yr is negligible nowadays.

Up to now we considered only the GA effect caused by the Galactocentric acceleration of the LSR due to the pure circular planar Galactic rotation.
This acceleration vector lies in the Galactic plane.
Some authors also discussed another possible acceleration of the Solar system directed in the perpendicular (vertical with respect to the Galactic plane)
direction.
\citet{Kopeikin2006} estimated a possible impact of the peculiar acceleration of the Solar system with respect to the LSR on GA magnitude.
They found that the residual proper motion is smaller in amplitude than 1~$\mu$as/yr.
\citet{Xu2012b} estimated the vertical component of the Solar system acceleration from VLBI observations
and obtained the magnitude of $3.95 \pm 0.47$~mm/s/yr ($2.7 \pm 0.3~\mu$as/yr) from global solution and $4.53 \pm 1.88$~mm/s/yr ($3.1 \pm 1.3~\mu$as/yr)
from the analysis of the solar velocity time series.
This result was not confirmed by \citet{Titov2013} who obtained the magnitude of the vertical component of
$0.3 \pm 1.3$~mm/s/yr ($0.2 \pm 0.9~\mu$as/yr) from DR solution and $-1.3 \pm 1.4$~mm/s/yr ($-0.9 \pm 1.0~\mu$as/yr) from DRQ solution.
One can see that these estimates of the vertical component of the Solar system acceleration are poorly consistent, which does not allow yet
to propose a practical model for reduction of the observation for this effect.

\subsection{Rotation to equatorial system}

For most of astrometric applications, after computation of the GA effect in Galactic coordinate system, it is converted to equatorial system,
which can be another source of error in GA modeling.
The rigorous algorithm for transformation of the source motion from Galactic to equatorial coordinate system is given by \citep{Murray1983}

\begin{equation}
\def\arraycolsep{1pt}
\begin{array}{l}
\mathbf{p}_1 = \left[ \begin{array}{c} -\sin l \\ \cos l \\ 0 \end{array} \right] , \quad
\mathbf{q}_1 = \left[ \begin{array}{c} -\sin b \cos l \\ -\sin b \sin l \\ \cos b \end{array} \right] , \\[5ex]
\mathbf{p}_2 = \left[ \begin{array}{c} -\sin \alpha \\ \cos \alpha \\ 0 \end{array} \right] \,, \quad
\mathbf{p}_G = \mathbf{G} \mathbf{p}_2 \,, \\[5ex]
\left[ \begin{array}{l} \mu_\alpha \cos\delta \\ \mu_\delta \end{array} \right]
=
\left[ \begin{array}{cc} \phantom{-} \mathbf{p}_G \mathbf{p}_1 \quad \mathbf{p}_G \mathbf{q}_1 \\
                                   - \mathbf{p}_G \mathbf{q}_1 \quad \mathbf{p}_G \mathbf{p}_1 \\
\end{array} \right]
\left[ \begin{array}{l} \mu_l \cos b \\ \mu_b \end{array} \right] \,,
\end{array}
\label{eq:gal2eq}
\end{equation}
where $\mathbf G$ is the rotation matrix between Galactic and equatorial coordinate systems.
Currently an inofficial standard of the transformation between Galactic and ICRS (equatorial) systems is defined by the matrix \citep{ESA1997}

\begin{equation}
\small
\def\arraycolsep{1pt}
\left[
\begin{array}{c}
-0.054875560416 \ -0.873437090235 \ -0.483835015549 \\
+0.494109427876 \ -0.444829629960 \ +0.746982244497 \\
-0.867666149019 \ -0.198076373431 \ +0.455983776175 \\
\end{array}
\right]
\label{matr:hipparcos}
\end{equation}

Alternative approaches to construction of transformation matrix based on the latest observations were proposed by \citet{Liu2011a,Liu2011b}.
Test computations have shown that the differences in proper motions obtained with several matrices discussed in these papers and
matrix \ref{matr:hipparcos} is less than 0.04~$\mu$as/yr.
Therefore, either matrix can be used to convert the GA-induced motions form Galactic to equatorial system.

\subsection{Modeling source position at given epoch}

The source position at epoch $t$ is computed from position at the catalog epoch $t_0$ and the source proper (apparent) motion $\mu$ by
\begin{equation}
\begin{array}{rcl}
\alpha(t) &=& \alpha(t_0) + \mu_\alpha (t-t_0) \,, \\
\delta(t) &=& \delta(t_0) + \mu_\delta (t-t_0) \,, \\
\end{array}
\label{eq:2epoch}
\end{equation}
where $\mu_\alpha$ and $\mu_\delta$ are computed by Eq~\ref{eq:gal2eq}.

Consider in detail the definition of $t_0$.
In fact, there is a choice of two possibilities.
First is that $t_0$ is taken to be equal to the mean epoch of the observation of given source.
In such a case the final catalog source position will be generally given at different epochs equal to mean epoch of observations of each source.
Such a catalog will formally have the source positions the same as a catalogs computed without GA modeling, but it will be not independent of the
epoch as currently supposed.
Using of such a catalog is not convenient and may be confusing.

\begin{figure}
\centering
\includegraphics[clip,width=0.48\textwidth]{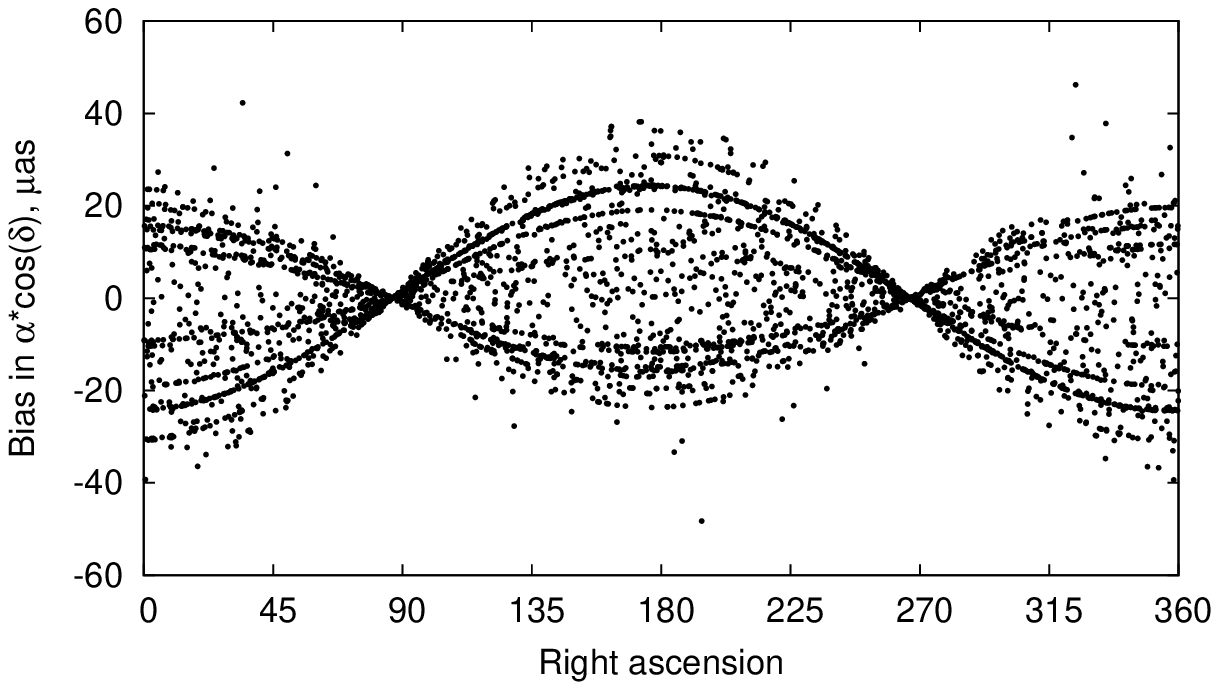}
\includegraphics[clip,width=0.48\textwidth]{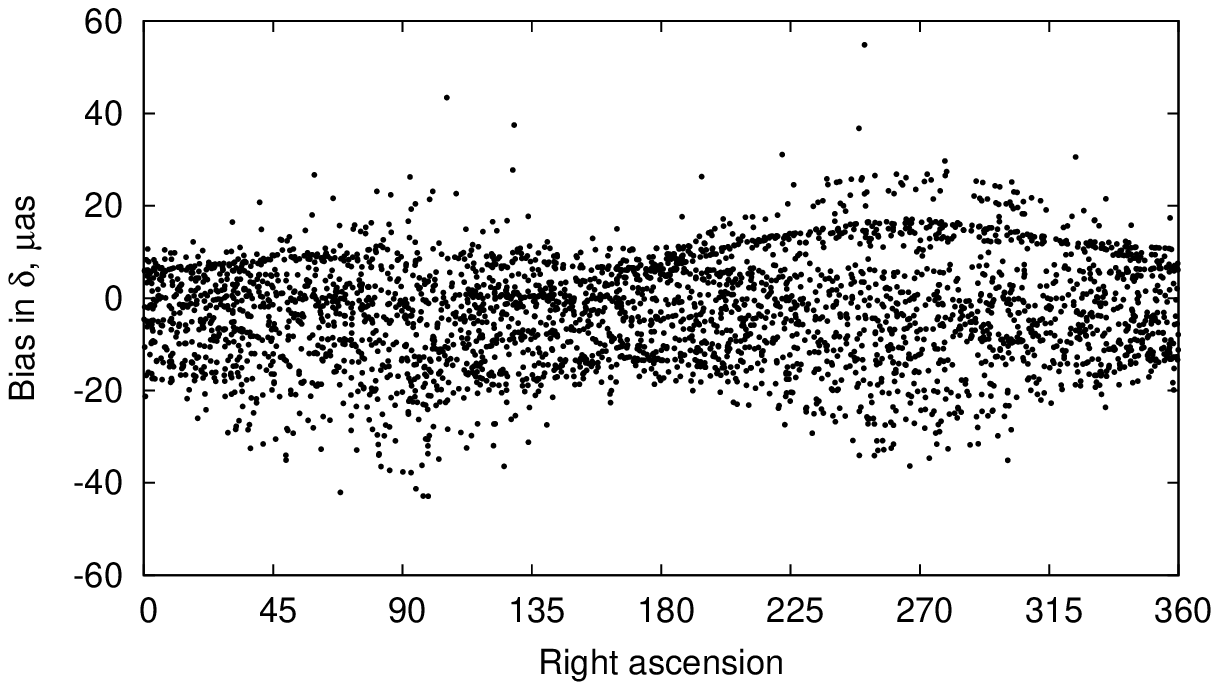}
\caption{Bias of source position between the current catalogs and catalogs computed with accounting for the GA at the epoch J2000.0.
The biases are computed using the actual mean epochs of the ICRF2 sources with respect to epoch J2000.0.}
\label{fig:bias}
\end{figure}

Traditional for astrometry is the choice of the same conventional epoch $t_0$ for all the catalog sources.
The epoch $t_0$ = J2000.0 is the most natural definition.
In such a case, after computation of new catalog with accounting for the GA we will have the bias in the source position depending on the source
coordinates and difference between the mean epoch of observations and $t_0$.
Figure~\ref{fig:bias} gives shows the value of the bias for all ICRF2 sources.

The pattern of Figure 3 reveals some interesting features.
The bias for given source depends on both the difference between the middle epoch of observations and source coordinates.
The sources located near right ascension 86$^\circ$ and 266$^\circ$, corresponding to directions to the Galactic center and anticenter,
has zero bias in this coordinate, as follows from Eq.~\ref{eq:galactic}, Fig.~\ref{fig:pm_aberr} and related discussion above.
Amplitude of the bias is quite substantial, especially for sources located near right ascension 176$^\circ$ and 356$^\circ$ and can exceed
40~$\mu$as.
It is explained by large deviation of the middle observation epoch from J2000.0, which varies for the ICRF2 sources from $-12.4$~yr to 9.0~yr.
Also, one can clearly see several curves composed of groups of VCS (VLBA calibrator survey) sources observed in the same epochs, see \citet{Ma2009p}
for details of the ICRF2 structure.

It is important that introducing such a bias is only a one-time inconvenience, which we inevitably have to come through to return to the traditional
and more rigorous astrometric practice.

%%%%%%%%%%%%%%%%%%%%%%%%%%%%%%%%%%%%%%%%%%%%%%%%%%%%%%%%%%%%%%%%%%%%%%%%%%%%%%%%%

\section{Conclusion}
\label{sect_conclusions}

The GA impact on the apparent motion of the celestial bodies is important for $\mu$as astrometry.
However, its implications should not be exaggerate.
It should be considered and treated just an astrometric reduction, one of many others.
The GA induced apparent motions of the extragalactic objects positions can be modelled with sufficient accuracy.
The most critical for the accuracy of the GA modeling is the accuracy of the GA constant $A$.
The value of $A = 5 \pm 0.3~\mu$as/yr based on the latest measurement of the Galactic rotation parameters
allows to account for at least 90\% of the full GA effect, which is sufficient for modern astrometry.
Indeed, the $A$ value will improve over time from new VLBI and space observations, as is the case with all the astronomical constant.

Finally, we can conclude that no new ICRS/ICRF concept is needed because of the GA effect.
We just have to return to traditional many-century astrometric practice of modeling the position of a celestial object as linear trend with the
two parameters, position at the initial (conventional) epoch and apparent (proper) motion.
Also we have to return to the traditional practice to publish, e.g., VLBI-based source position catalogs at standard epoch, preferably J2000.0,
to which the astrometric equations are mostly referred.
It looks mandatory to move to this strategy for the next VLBI-based ICRS realization ICRF3 to achieve the $\mu$as level of accuracy \citep{Jacobs2013jsr}.

Another consequence of moving to the classical paradigm is that mutual orientation between two CRF realization is no longer epoch-independent,
but is referred to the selected conventional epoch, preferably J2000.0.
Indeed, both compared catalogs should be brought to this epoch.
In particular, this strategy should be employed for alignment of the Gaia Celestial Reference Frame (GCRF) to ICRF.

\section*{Acknowledgements}
The author is grateful to the GSFC VLBI group for making their series of astrometric catalogs available.
Constructive comments and suggestions from the anonymous referee are gratefully acknowledged.

\bibliography{my_eng,astronomy,geodesy,math,galaxy}
\bibliographystyle{mn}

%%%%%%%%%%%%%%%%%%%%%%%%%%%%%%%%%%%%%%%%%  ERRATUM  %%%%%%%%%%%%%%%%%%%%%%%%%%%%%%%%%%%%%%%%%%%%%%%%%

\clearpage

\setcounter{equation}{0}

~\vskip 2em
\noindent
{\LARGE \bf Erratum}

\vskip 2em
This is an erratum to the paper ``On the implications of the Galactic aberration in proper motions for celestial reference frame''
published in MNRAS 445, 845 (2014).
As was pointed out by Patrick~Wallace (private communication), the text related to matrix (5) is, strictly speaking, not fully accurate
and needs clarification.

It was stated in the original paper that the matrix (5) is taken from Perryman \& ESA (1997).
In fact, Perryman \& ESA (1997), Eqs. (1.5.9) and (1.5.10), defined as the primary quantities three orientation angles:

\noindent right ascension of the north galactic pole in ICRS:

$\alpha_G = 192\hspace{0.1ex}.\hspace{-0.75ex}^{\circ}85948$,

\noindent declination of the north galactic pole in ICRS:

$\delta_G =  27\hspace{0.1ex}.\hspace{-0.75ex}^{\circ}12825$, and

\noindent galactic longitude of the ascending node of the galactic plane on the
equator of ICRS:

$l_{\Omega} = 32\hspace{0.1ex}.\hspace{-0.75ex}^{\circ}93192$.

Perryman \& ESA (1997) wrote: ``The angles $\alpha_G$, $\delta_G$ and $l_{\Omega}$ are to be regarded as exact quantities. From them, the
transformation matrix may be computed to any desired accuracy''.
They provided a transformation matrix computed with the above angles to ten decimal
places (transposed for compatibility with our paper):
\begin{equation}
\left[
\begin{array}{c}
-0.0548755604 \ -0.8734370902 \ -0.4838350155 \\
+0.4941094279 \ -0.4448296300 \ +0.7469822445 \\
-0.8676661490 \ -0.1980763734 \ +0.4559837762 \\
\end{array}
\right] .
\end{equation}

Matrix (5) in our original paper:
\begin{equation}
\def\arraycolsep{1pt}
\left[
\begin{array}{c}
-0.054875560416 \ -0.873437090235 \ -0.483835015549 \\
+0.494109427876 \ -0.444829629960 \ +0.746982244497 \\
-0.867666149019 \ -0.198076373431 \ +0.455983776175 \\
\end{array}
\right]
\end{equation}
was computed using the same orientation angles with higher precision
to provide microarcsecond levels of accuracy for coordinate transformation.
So, not this matrix itself, but the orientation angles between the galactic coordinate system and ICRS used for the matrix computation were
taken from Perryman \& ESA (1997).

A separate question is whether the orientation angles can be taken as constants independent of the epoch.
That is true in the case of Galactic rotation without precession, which, evidently, cannot currently be verified.

The results and conclusions of the original paper are not affected by this discussion.

\vskip 2em
\noindent{\bf REFERENCES}

\medskip

{
\leftskip=2ex
\parindent=-1ex

Perryman M. A. C., ESA eds., 1997, ESA SP-1200, The HIPPARCOS
and TYCHO catalogues. Astrometric and photometric star catalogues
derived from the ESA HIPPARCOS Space Astrometry Mission.
ESA Publications Division, ESTEC, Noordwijk, The Netherlands

}

\end{document}